\documentclass[useAMS,usenatbib]{mn2e}
\usepackage{graphicx}
\usepackage{color}

\title[Two types of glitches in a solid quark star model]{Two types of glitches in a solid quark star model}
\author[E. P. Zhou, J. G. Lu, H. Tong and R. X. Xu]{E. P. Zhou$^{1}$, J. G. Lu$^{1}$, H. Tong$^{2}$ and R. X. Xu$^{1,3}$\thanks{E-mail:zhouenpingz715@sina.com(EPZ);r.x.xu@pku.edu.cn(RXX)}\\
$^{1}$State Key Laboratory of Nuclear Science and Technology and School of Physics, Peking University, Beijing 100871, P. R. China\\
$^{2}$Xinjiang Astronomical Observatory, Chinese Academy of Sciences, Urumuqi, Xinjiang, 830011, China\\
$^{3}$Kavil Institute for Astronomy and Astrophysics, Peking University, Beijing 100871, P. R. China}
\begin{document}


\pagerange{\pageref{firstpage}--\pageref{lastpage}} \pubyear{2002}

\maketitle

\label{firstpage}

\begin{abstract}
Glitch (sudden spinup) is a common phenomenon in pulsar observations. However, the physical mechanism of glitch is still a matter of debate because it depends on the puzzle of pulsar's inner structure, i.e., the equation of state of dense matter. Some pulsars (e.g., Vela-like) show large glitches ($\Delta \nu/\nu \sim 10^{-6}$) but release negligible energy, whereas the large glitches of AXPs/SGRs (anomalous X-ray pulsars/soft gamma repeaters) are usually (but not always) accompanied with detectable energy releases manifesting as X-ray bursts or outbursts.
We try to understand this aspect of glitches in a starquake model of solid quark stars.
There are two kinds of glitches in this scenario: bulk-invariable (Type I) and bulk-variable (Type II) ones. The total stellar volume changes (and then energy releases) significantly for the latter but not for the former.
Therefore, glitches accompanied with X-ray bursts (e.g., that of AXP/SGRs) could originate from Type II starquakes induced probably by accretion, while the others without evident energy release (e.g., that of Vela pulsar) would be the result of Type I starquakes due to, simply, a change of stellar ellipticity.
\end{abstract}

\begin{keywords}
 dense matter - pulsars: general - stars: magnetar - stars: neutron.
\end{keywords}

\section{Introduction}




Pulsars are accurate clocks in the Universe. During pulsar timing studies, glitches are discovered~\citep{r}.
A glitch is a sudden increase in pulsar's spin frequency, $\nu$, and the observed fractions $\Delta \nu/\nu$ range
between $10^{-10}$ and $10^{-5}$, the distribution of which is bimodal with peaks at approximately $10^{-9}$ and
$10^{-6}$ (Yu et al. 2013).
It can help us understand the inner structure of pulsars.
It has been more than 40 years since the discovery of Vela pulsar glitch and a lot of studies
on its physical origin have been carried out since then. In neutron star models, a pulsar is thought to be a fluid star with
a thin solid shell. The physical mechanism behind glitches is believed to be the coupling and decoupling between outer crust
(rotating slower) and the inner superfluid (rotating faster) \citep{a2,a1}. However, the absence of evident energy
release during even the largest glitches ($\Delta \nu/\nu \sim 10^{-6}$) of Vela pulsar is a great
challenge to this glitch scenario \citep{g2,h}. The glitches detected from AXPs/SGRs
(anomalous X-ray pulsars/soft gamma repeaters),
usually accompanied with energy release though the maximum amplitude of which is also $\Delta \nu/\nu \sim 10^{-6}$, represent
an additional challenge to the glitch scenario in neutron
star models~\citep{k,d2,t}.

In spite of these problems, however, it is worth noting that the glitch mechanism depends on
the state of cold matter at supra-nuclear density, the solution
of which is relevant to the challenging problem of particle physics, the non-perturbative
quantum chromodynamics.
Nevertheless, great efforts have been tried to model the inner structure of pulsars. Traditionally speaking,
quarks are confined in hadrons of neutron stars, while a quark star is composed of de-confined quarks.
While a solid quark star is a condensed object of quark
clusters, which distinguishes from conventional both neutron and quark stars~\citep{x1,xu2010,xu2013}.
The solid quark star (i.e. quark cluster star) is quite different from the traditional quark stars.
The properties which are common in traditional quark stars, e.g. colour superconductivity with colour-flavor locking~\citep{o},
are not expected in solid quark stars as the quarks in such stars cannot be treated as
free fermion gas any more. The magnetic field of a solid quark star will also be quite different~\citep{x2005} from that of a
traditional quark star~\citep{i} because of the different magnetic origins~\citep{c2007,x2005}.
The equation of state is very stiff in the solid quark star model, which is favored by the
discovery of massive pulsars~\citep{l1,d,l2,a}.
A special kind of quark-cluster, H-cluster, has also been considered~\citep{l3}.
Additionally, the peculiar X-ray flare and the plateau of $\gamma$-ray burst could be relevant
to a solid state of quark matter (Xu \& Liang 2009; Dai et al. 2011).
Glitches are thought to result from starquakes in a solid star model~\citep{b}. The general glitch behaviors such as the
amplitude and the time interval could be well
reproduced by parameterizing shear modulus and critical
stress. The post-glitch behavior could also be explained as the damped oscillation of the
solid quark star~\citep{z1}.
This glitch model has also been extended to explain the timing behaviors of slow glitches~\citep{p}.

There are two kinds of starquakes in a solid quark star model: bulk-invariable and
bulk-variable ones~\citep{p}. We call the former Type I and the latter Type II starquakes.
Two types of starquakes will result in two types of glitches, respectively.
On one hand, as a pulsar spins down, the ellipticity would decrease gradually to maintain
the equilibrium configuration if the star is in a fluid state. However, for a solid quark star, elastic
energy will accumulate to resist the change in shape. When this elastic energy exceeds a
critical value that the star can no longer stand against, a bulk-invariable starquake occurs
({\em Type I}).
On the other hand, even without rotation, a solid star may shrink its volume abruptly,
especially in case of accretion which can cause substantial mass
and gravity gain, and a bulk-variable starquake happens ({\em Type II}).
A real glitch could be a mixture of these two, but may be dominated by either.

Our calculations find significant energy releases
for the bulk-variable starquakes, but not for the bulk-invariable ones.
Therefore it is suggested that Type I and Type II starquakes result in Type I glitches (glitches in normal
pulsars and some glitches in AXP/SGRs, which are radiation quiet) and Type II glitches (some glitches on AXP/SGRs, which are accompanied with radiative anomalies), respectively.
In this regime, X-ray burst could be detected after a Type-II glitch, however one could not discover X-ray enhancement after a Type-I glitch.

\section[]{Two types of starquakes and corresponding energy releases}

Vela-like glitches are assumed to be Type I glitches since they are discovered earlier.
However, in our model, it is easier to figure out the energy release during a Type II glitch.
Considering this, Type II glitches will be discussed first.

\subsection{Bulk-variable (Type II) starquake}

For a quark star with relatively low mass ($M<1.0 M_\odot$), it is self-bound by
strong interaction and gravity can be neglected. This results in an approximately $M\sim R^{3}$ relation for
lower mass quark stars. While the relation is violated for stars with larger mass ($M>1.0 M_\odot$)
since the gravity dominates instead of strong interaction. The $M$-$R$ relation for a pure
gravity-bound star is $M\sim R^{-3}$. This indicates the existence of a maximum radius in the
$M$-$R$ relation of quark stars \citep{x2}. Many $M$-$R$ relations of quark stars also prove
the fact that there should be a maximum radius \citep{l1,g1}.
The mass of a quark star may exceed that corresponds to the maximum radius by accretion. In this case, the radius
of the star would decrease but should still be larger than the equilibrium radius given by the $M$-$R$ relation
because elastic energy will be accumulated to resist the change in configuration.
As the accretion continues, the elastic energy will finally exceed the limitation of the solid
structure, resulting in a starquake which makes the star entirely collapse to reach
the supposed stable radius. We can describe this kind of glitch as a global reduce in radius ($\delta$$R$).

Supposing that the pulsar is a solid quark star with mass $M$, radius $R$ and angular spin
velocity $\Omega$ before the starquake, the total energy before the starquake is
\begin{equation}
E_{\mathrm{total}}=E_{\mathrm{k}}+E_{\mathrm{g}}=\frac{I\Omega^2}{2}-\frac{3GM^2}{5R},
\label{Eq.1}
\end{equation}
in which $I$ is moment of inertia of the pulsar and $G$ is the gravitational constant.
As the duration of the glitch is quite short, the conservation of angular momentum can be
applied. The total energy can be written as
\begin{equation}
E_{\mathrm{total}}=\frac{L^2}{2I}-\frac{3GM^2}{5R},
\label{Eq.2}
\end{equation}
where $L$ is the angular momentum.

After the starquake, the change in total energy is
\begin{equation}
\delta E=\frac{3GM^2}{5R}\frac{\delta R}{R}-\frac{L^2}{2I}\frac{\delta I}{I},
\label{Eq.3}
\end{equation}
in which the moment of inertia can be taken as that of a spherical star with mass $M$ and radius $R$:
\begin{equation}
I=\frac{2}{5}MR^2,
\label{Eq.4}
\end{equation}
indicating that
\begin{equation}
\frac{\delta I}{I}=\frac{2\delta R}{R}.
\label{Eq.5}
\end{equation}
Therefore the change of total energy is:
\begin{equation}
\delta E=(\frac{3GM^2}{5R}-\frac{L^2}{I})\frac{\delta R}{R}\approx\frac{3GM^2}{5R}\frac{\delta R}{R}.
\label{Eq.6}
\end{equation}
For a pulsar with 1.4$\,M_\odot$, 10$\,$km and rotation period larger than 1$\,$ms, the approximation of
\begin{equation}
\frac{3GM^2}{5R}\gg\frac{L^2}{I},
\label{Eq.8}
\end{equation}
can be applied.

In this case, the amplitude of the glitch can be given by the conservation of angular momentum
\begin{equation}
\delta L=\delta I \Omega+I \delta \Omega=0,
\label{Eq.9}
\end{equation}
leading to
\begin{equation}
\frac{\delta \Omega}{\Omega}=-\frac{\delta I}{I}=-\frac{2\delta R}{R}.
\label{Eq.10}
\end{equation}
Observationally, the amplitude of a glitch is
\begin{equation}
\frac{\delta \nu}{\nu}=\frac{\delta \Omega}{\Omega}=-\frac{2\delta R}{R}.
\label{Eq.36}
\end{equation}
Thus a shrinkage ($\delta R<0$) of the star results in a spin-up glitch.

The change of total energy is
\begin{equation}
\delta E=\frac{3GM^2}{5R}\frac{\delta R}{R}=-\frac{3GM^2}{10R}\frac{\delta \nu}{\nu}.
\label{Eq.37}
\end{equation}
Consequently the energy release during a Type II starquake is
\begin{equation}
|\delta E|=\frac{3GM^2}{10R}\frac{\delta \nu}{\nu}\sim10^{47}\,{\mathrm{erg}}(\frac{M}{1.4\,M_{\odot}})^{2}(\frac{R}{10^6\,{\mathrm{cm}}})^{-1}({\frac{\delta \nu}{\nu}}/{10^{-6}}).
\label{Eq.35}
\end{equation}

This theoretical energy release is sufficient to explain the outbursts of AXPs/SGRs which are thought to be related to
glitches. For instance, the typical fraction of glitches on 1E2259 is $\delta \nu/\nu\sim10^{-6}$. So the resulting
energy release is large enough to understand the
corresponding bursts with energy release of $10^{40}\,{\mathrm{erg}}$ \citep{w2}.

\subsection{Bulk-invariable (Type I) starquake}

\begin{figure}
\includegraphics[width=0.35\textwidth]{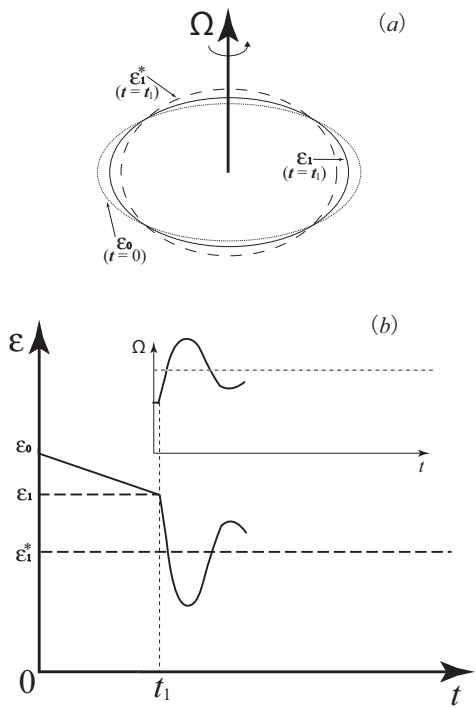}
\caption{An illustration of the variable $\varepsilon$. The ellipticity of the star is exaggerated in this figure.
The value of $\varepsilon_0$ is the non-elastic-energy ellipticity (the ellipticity when the star became a solid).
As the star spins down, the ellipticity of a Maclaurin ellipsoid becomes ${\varepsilon_1}^{*}$.
However, as a solid star, its real ellipticity ($\varepsilon_1$) can't reach this value. Part b shows the change
of $\varepsilon$ parameter corresponding to the change of $\Omega$ during the glitch. A similar figure was
first used by Peng \& Xu (2008) as an illustration.}
\label{fig:fit1}
\end{figure}

For pulsars of which the accretion can be
neglected, starquakes are supposed to happen in another way.
It is known that the equilibrium configuration of a rotating star is a ellipsoid instead of a
perfect sphere. There are also many models describing the deformation of a rotating fluid
star. The Maclaurin ellipsoid is one of the most widely used models.
In this model, the star is suggested to be a incompressible fluid. The ellipticity of a star
with density $\rho$ relies on the the angular spin velocity of the
star. The analytical relationship is \citep{c}
\begin{equation}
\Omega^2=2\pi G\rho[\frac{\sqrt{1-e^2}}{e^3}(3-2e^2)\sin^{-1}e-\frac{3(1-e^2)}{e^2}].
\label{Eq.11}
\end{equation}
The ellipticity $e$ is given by
\begin{equation}
e=\sqrt{1-\frac{c^2}{a^2}},
\label{Eq.12}
\end{equation}
where $a$ and $c$ are the semimajor and semiminor axes, respectively.
An approximation of
\begin{equation}
\Omega=2e\sqrt{\frac{2\pi G \rho}{15}}
\label{Eq.38}
\end{equation}
can be made when $e$ is small (i.e., for slow rotators).

However, the relationship is useful only for fluid stars. For solid stars such as quark
cluster stars, the elastic energy increases as the star resists the change in shape.
For solid stars, the total energy of the star is \citep{b}
\begin{equation}
E_{\mathrm{total}}=E_{\mathrm{k}}+E_{\mathrm{g}}+E_{\mathrm{ela}}=E_0+\frac{L^2}{2I}+A\varepsilon^2+B(\varepsilon-\varepsilon_0)^2,
\label{Eq.13}
\end{equation}
where
\begin{equation}
A=\frac{3GM^2}{25R},
\label{Eq.14}
\end{equation}

\begin{equation}
B=\frac{\mu V}{2},
\label{Eq.15}
\end{equation}
The term of $A\varepsilon^2$ represents the difference between the gravitational energy of the ellipsoid and that of a spheriod with same mass and density ($E_{0}$) and the term of $B(\varepsilon-\varepsilon_0)^2$ represents the elastic energy. And the kinetic energy is written is the form of $L^{2}/2I$.
The variable $\varepsilon$ is a reduced ellipticity:
\begin{equation}
\varepsilon=\frac{I-I_0}{I_0},
\label{Eq.16}
\end{equation}
where $I_0$ is the moment of inertia of a spherical star with
same mass and density and $I$ is the real moment of inertia considering the deformation
by rotation. When the ellipticity is small, we can take a simplification of
\begin{equation}
\varepsilon=\frac{1}{3}e^2.
\label{Eq.17}
\end{equation}
A proper approximation for a pulsar with period of 10ms is $\varepsilon\sim 10^{-3}$.
And $\varepsilon_0$ is a critical ellipticity of the star, at which ellipticity the elastic
energy is zero. It can be taken as the ellipticity at the end of the previous glitch if we suggest that
all the elastic energy is released in a glitch.
Generally speaking, a newly born quark star in a core collapse supernova is hot and
can be treated as a fluid star. Thus when it loses rotation energy, the ellipticity also
decreases as a Maclaurin ellipsoid. But when it cools down to certain temperature, the
elastic energy comes to exist and it can no longer be treated as fluid.
When the pulsar continues spinning down, its ellipticity will decrease less than that of a Maclaurin
ellipsoid. In this process, the elastic energy is accumulated. When the elastic energy is
large enough, a starquake may happen and in this short duration the star can be treated as
fluid again~\citep{p}. Hence there will be a sudden decrease in $\varepsilon$. From the definition
of the $\varepsilon$ we know this means a sudden decrease in moment of inertia. Therefore there
will be a increase in the angular spin velocity, given by
\begin{equation}
\frac{\delta \Omega}{\Omega}=-\frac{\delta I}{I}=-\frac{\delta \varepsilon}{1+\varepsilon}.
\label{Eq.18}
\end{equation}
Considering
\begin{equation}
\delta \varepsilon \ll \varepsilon \ll 1,
\label{Eq.19}
\end{equation}
we can obtain
\begin{equation}
\frac{\delta \Omega}{\Omega}\sim -\delta \varepsilon.
\label{Eq.20}
\end{equation}
This is the amplitude of a glitch resulted from the bulk-invariable starquake.

A Type I starquake consists of 3 steps as shown in Fig.1: 1) a normal spin down phase which begins
at the end of last glitch. The elastic energy is accumulated during this phase; 2) the glitch epoch. At this
time the star loses its elastic energy and can be seen as a fluid star; 3) the glitch phase. During this phase the star
changes its shape and sets up a new equilibrium at the end of this phase.
In the normal spin down phase, the ellipticity of the star should be stable and change
smoothly as the spin frequency changes. For a given angular momentum, the condition of equilibrium can be given as
\begin{equation}
\frac{\partial E}{\partial \varepsilon}=0.
\label{Eq.21}
\end{equation}
This condition can also be applied by the end of the glitch because the equilibrium is set up again.

Right before the glitch epoch, the total energy of the star is,
\begin{equation}
E_{t_{1}-0}=E_{\mathrm{k}}+E_{\mathrm{g}}+E_{\mathrm{ela}}=E_0+\frac{L^2}{2I_{0}(1+\varepsilon_{1})}+A\varepsilon_{1}^2+B(\varepsilon_{1}-\varepsilon_0)^2.
\label{Eq.30}
\end{equation}
While right after the glitch epoch, the total energy of the star becomes
\begin{equation}
E_{t_{1}+0}=E_{\mathrm{k}}+E_{\mathrm{g}}=E_0+\frac{L^2}{2I_{0}(1+\varepsilon_{1})}+A\varepsilon_{1}^2.
\label{Eq.31}
\end{equation}
At the end of the glitch, the energy is
\begin{equation}
E_{\mathrm{final}}=E_{\mathrm{k}}+E_{\mathrm{g}}=E_0+\frac{L^2}{2I_{0}(1+\varepsilon_{1}^{*})}+A(\varepsilon_{1}^{*})^2.
\label{Eq.32}
\end{equation}

Considering that the ellipticity change is quite small during the glitch, the condition of equilibrium at the end of the glitch, as well as the possible damped vibrations~\citep{z1} one can obtain
\begin{equation}
E_{\mathrm{final}}\approx E_{t_{1}+0}.
\label{Eq.33}
\end{equation}
Thus the total energy change during the glitch is the elastic energy.

Choosing two epoches in the normal spin down phase: $t_{0}$ and $t_{1}$, one with angular momentum $L_0$,
ellipticity $\varepsilon_0$ and the other $L_1$, $\varepsilon_1$, we have
\begin{equation}
\label{Eq.22}
\frac{\partial E}{\partial \varepsilon}\mid_{L_0}=-\frac{{L_0}^2}{2I_0(1+\varepsilon_0)^2}+2A\varepsilon_0+2B(\varepsilon_0-\varepsilon_0)=0,
\end{equation}

\begin{equation}
\frac{\partial E}{\partial \varepsilon}\mid_{L_1}=-\frac{{L_1}^2}{2I_0(1+\varepsilon_1)^2}+2A\varepsilon_1+2B(\varepsilon_1-\varepsilon_0)=0.
\label{Eq.23}
\end{equation}
Combining Equation (\ref{Eq.22}) and Equation (\ref{Eq.23}), we obtain

\begin{equation}
\varepsilon_0-\varepsilon_1=\frac{1}{4(A+B)}I_0({\Omega_0}^2-{\Omega_1}^2).
\label{Eq.25}
\end{equation}
Thus one can have
\begin{equation}
E_{\mathrm{ela}}=B(\varepsilon_1-\varepsilon_0)^2=\frac{B}{2(A+B)}\frac{1}{2}I_0({\Omega_0}^2-{\Omega_1}^2)(\varepsilon_0-\varepsilon_1).
\label{Eq.26}
\end{equation}
Then we can obtain
\begin{equation}
E_{\mathrm{ela}}<\frac{B}{2(A+B)}|\delta E_{\mathrm{k}}|(\varepsilon_0-\varepsilon_1).
\label{Eq.27}
\end{equation}
This is also the energy that will be released in the glitch.
According to the calculations by Zhou et al. (2004), the change rate of ellipticity during the normal spin down phase is
\begin{equation}
\dot{\varepsilon}=\frac{I_{0}\Omega\dot{\Omega}}{2(A+B)}.
\label{Eq.34}
\end{equation}
While the change rate of the supposed equilibrium configuration of a Maclaurin sphere is
\begin{equation}
\dot{\varepsilon}_{\mathrm{Mac}}=\frac{I_{0}\Omega\dot{\Omega}}{2A}.
\label{Eq.39}
\end{equation}
It's natural to consider that the difference between the real configuration and the Maclaurin equilibrium configuration is eliminated during the glitch. Thus we can work out the ratio of the ellipticity change during the normal spin down phase ($\varepsilon_1-\varepsilon_0$) and that during the glitch ($-\delta \Omega/\Omega$) as
\begin{equation}
\varepsilon_1-\varepsilon_0=-\frac{A}{B}\frac{\delta \Omega}{\Omega}.
\label{Eq.40}
\end{equation}
Applying this ratio into Equation (\ref{Eq.27}), the total energy release is
\begin{equation}
\delta E\sim\frac{A}{2(A+B)}|\delta E_{\mathrm{k}}|\frac{\delta \Omega}{\Omega}.
\label{Eq.41}
\end{equation}
Consider that $B$ is smaller than $A$ \citep{z1}, the total energy release is
\begin{equation}
\delta E\sim\frac{1}{2}|I\Omega\dot{\Omega}|t\frac{\delta \Omega}{\Omega}=\frac{1}{2} E_{\mathrm{k}}\frac{t}{\tau_{\mathrm{c}}}\frac{\delta \Omega}{\Omega},
\label{Eq.42}
\end{equation}
in which $t$ is the interval between two glitches ($t\sim10^{6}\,\mathrm{s}$ for Vela) and $\tau_{\mathrm{c}}$ ($\mathbf{\tau_{\mathrm{c}}\sim P/2\dot{P}\sim-\Omega/2\dot{\Omega}}$) is the
characteristic age of a pulsar.

According to the observations~\citep{d2002}, the period and period derivative of Vela are 0.089\,s and 1.25$\times10^{-13}$, respectively. Assuming that Vela is a pulsar with mass of 1.4\,$M_{\odot}$ and radius of 10\,km, the total energy released during a Type I glitch
can be written as
\begin{equation}
\delta E\sim4\times10^{36}\,{\mathrm{erg}}(\frac{t}{10^{6}\,{\mathrm{s}}})(\frac{\delta\nu}{\nu}/{10^{-6}}).
\label{Eq.44}
\end{equation}
Helfand et al. (2001) have made a constraint on the X-ray flux enhancement of Vela 35 days after a $\Delta \nu/\nu \sim 10^{-6}$
glitch, the upper limit of which is 1.2$\times10^{30}$\,erg$\,\mathrm{s}^{-1}$. According to our result, even if we assume that all the energy release (4$\times10^{36}$\,erg) during the glitch is radiated as X-ray photons and the flux keeps constant during the 35 days (3$\times10^{6}$\,s), the resulting flux (1.3$\times10^{30}$\,erg$\,\mathrm{s}^{-1}$) is somehow consistent with their observational constraints. In fact, not all the energy would be converted into X-ray radiation and it's likely that the flux decreases with time.

\begin{figure}
\includegraphics[width=0.5\textwidth]{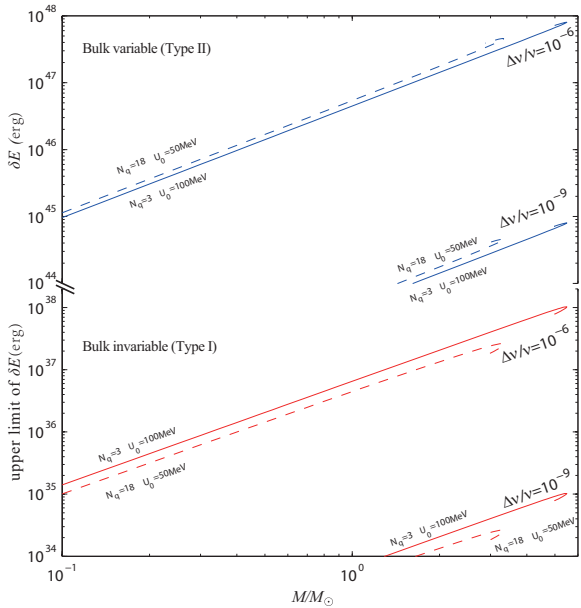}
\caption{The total energy release during the bulk-variable (Type II) glitches
and bulk-invariable (Type I) glitches with amplitudes of $10^{-6}$ and $10^{-9}$.
For the Type I case, only the upper limit of energy release is shown and
the real energy release will be reduced by a factor of $A/[2(A+B)]$ (see Equation (\ref{Eq.41}) please).
The Lennard-Jones interaction is applied as an approximation to work out the mass-radius relation \citep{l1}.
There are two main factors in this approximation: the number of quarks in one cluster ($N_{\mathrm{q}}$) and
the depth of the potential ($U_{0}$). The case of 3-quark clusters with potential of 100MeV (solid lines) and
18-quark clusters with potential of 50MeV (dashed lines) are considered. It is also worth noting that the
energy release during a Type I glitch is related to the time intervals between two glitches.
In this calculation the glitch is thought to happen once per month and the spin down power is calculated
according to the observational data of Vela.}
\label{fig:fit1s}
\end{figure}

\section{Detailed Calculations}

We have already figured out the theoretical energy releases with respect to the amplitudes of glitches
for pulsars with certain mass (1.4\,$M_{\odot}$) and radius (10\,km).
However, the equation of state is also an important factor besides all parameters mentioned above.
It directly influences the gravitational energy and the moment of inertia of the pulsar.

A detailed result for one specific equation of state can be seen in Fig.2.
The upper panel and lower panel show the energy release in
bulk-variable cases and an upper limit of the energy release in bulk-invariable cases, respectively.
The mass-radius relation
approached by a Lennard-Jones interaction approximation is used in this work \citep{l1}.
For the bulk-invariable case, the period and period derivative are set to fit the
observational data of Vela and the glitch interval is one month.

In the calculation of the Type I glitches, we suggest that the ellipticity of pulsar reaches the critical Maclaurin
ellipticity ($\varepsilon_0$) at the end of each glitch, which means all the elastic energy is released during the
glitch. However, there could be some cases that only part of the elastic energy is released in a glitch \citep{p}, i.e.,
maybe only the surface of the star breaks up and changes its shape. So our result is only an upper limit of real
energy release during a Type I glitch.

It is clear that even the largest bulk-invariable glitch ($\delta\nu/\nu\sim10^{-6}$)
releases no more energy than $10^{38}$\,erg. While a $\delta\nu/\nu\sim10^{-9}$ bulk-variable
glitch is six orders of magnitude more energetic.

\section{discussions and conclusions}

The reason why the energy release during a Type I and a Type II starquake are
different is the fact that the matter near the equator contributes to most of the
moment of inertia.
A global collapse happens during a Type II starquake. The matter near the polar region
can hardly reduce the pulsar's moment of inertia when collapse. Actually, most of collapsed
matter could not contribute to the spinup of the pulsar when it releases gravitational
energy. Therefore, a bulk-variable Type-II starquake seems more energetic.
However, for a Type I starquake, what really changes is the matter distribution.
A Type I starquake could be regarded as a transport of matter from equator to the pole.
It is true that gravitational energy can also release because the radius at the polar region is
less than that at the equator, but reducing moment of inertia is much more efficient.
This is the reason why less energy releases but the star spins up a lot during a Type-I starquake.
Additionally, it has been discussed that the timescale for the elastic energy built up in a Type-I
starquake is longer than that in a Type II starquake. And the quake may happen in different
position in the star. These will also lead to quite different manifestations of energy release ~\citep{t}.
Observationally, glitches have been classified into two sorts according to the radiative properties which
is quite similar to what we have done in this paper ~\citep{t1}. The two types of starquakes in a solid quark
star model discussed above can account for the two types of glitches in observation.

As with the trigger of the starquake, we think that accretion should be the key factor for a Type II glitch.
As mentioned above, elastic energy develops substantially when a solid star gains
mass and thus gravity~\citep{x2}.
We may expect that Type II glitches are most likely to happen in AXP/SGRs since (1) they are
spinning slowly and (2) accretion would be possible there~\citep{chn} and observation
hints the existence of disk~\citep{w1}. It's also consistent with the 'quark star/fallback disk'
model in which AXP/SGRs are thought as solid quark stars surrounded by fallback disks~\citep{x2,t}.
However, for normal pulsars, such as Vela/Crab pulsars, the compact objects rotates
relatively faster, and the ellipticity change should be considerably important
during the evolution.
It is also worth noting that Type-II glitches could occur not only on solid quark stars with
large masses. Because of gravity, real stellar radius is always smaller
than that given by $M\sim R^3$ law. Certainly elastic energy is accumulated whenever the
accretion happens.
Another factor of accumulating anisotropic pressure distributed inside solid matter could be the temperature effect~\citep{p}.

For previous starquake-induced glitch models, it is known that large glitches on Vela pulsar
happen so frequently that the stress built up in the star is smaller than required. However, in our
model, large glitches on Vela do not necessarily imply that large amount of energy is accumulated.
What really matters is the initial ellipticity of Vela (i.e. the ellipticity when Vela became solid).
By suggesting the rotation period of Vela was 4\,ms when it solidified, the initial reduced ellipticity ($\varepsilon$$\sim$0.01) would be
large enough for Vela to suffer more than 10000 glitches with $\Delta \nu/\nu\sim 10^{-6}$ during its lifetime. It may infer a short timescale before the solidification of the newly born pulsar.
According to the theoretical conditions and observational hints~\citep{dai,x1}, a quark cluster star with the density of
two times nuclear density is most likely to solidify at the temperature of $T\sim0.5$\,MeV. It happens at about
1000 seconds after the formation of the star, which is reasonably short.
We also need a large constant of $B$ ($B\sim A$) in this model so that considerable part of ellipticity decrease happens during the glitch.

A real glitch may consist of both types, which means when the radius of an AXP/SGR shrinks, there could also
be a trend that part of the matter flow to the polar region. This could be a reason why the observed energy of the outbursts
is much less than that we predict for a Type II glitch. Another reason is that the majority of the energy release is
taken away by neutrino emission. The energy loss due to gravitational radiation depends on the
detailed behavior of the stellar oscillation during and after the glitch.
However, gravitational radiation is expected to be weak for a Type I glitch
because the total energy release is negligible.

In conclusion, it is found that two types of starquakes could occur in a solid quark star as it
evolves: Type I (bulk-invariable) and Type II (bulk-variable).
The total stellar volume decreases abruptly during a Type II starquake, but it is
conserved for Type I even if stellar elipticity changes discontinuously.
Consequently a pulsar may spin up suddenly, observed as a glitch, and it is then evident
that there are two types of glitches caused by each type of starquake in a solid quark star model.
A Type II glitch could be energetic enough for us to detect X-ray emission even if the
glitch amplitude of $\Delta \nu/\nu$ would be as small as $10^{-9}$. For a
Type I glitch, no X-ray enhancement could be detected even for a large glitch of
$\Delta \nu/\nu\sim 10^{-6}$.

\section*{Acknowledgments}

We would like to thank the pulsar group of PKU for useful discussions.
This work is supported by National Basic Research Program of China (2012CB821800), National
Natural Science Foundation of China (11225314 \& 11103021) and XTP project XDA04060604.

\label{lastpage}

\end{document}